# Ionic Current Rectification in Nanopores: Effects of Nanopore Material, Electrolyte and Surface Treatment

Andrea Doricchi[a,b±], Anastasiia Sapunova[a,c±], Ali Douaki[a,d], German Lanzavecchia[a,e], Shukun Weng[a,c], Mostafa Salehirozveh[f], Makusu Tsutsui[g], Mauro Chinappi[h], Dmitry Momotenko[i], Roman Krahne[a], and Denis Garoli*[a, l]

[a] Istituto Italiano di Tecnologia, Via Morego 30, 16136 Genova, Italy

[b] Università degli Studi di Genova, Dipartimento di Chimica e Chimica Industriale, Via Dodecaneso 31, 16146, Genova, Italy

[c] Università degli Studi di Milano-Bicocca, Dipartimento di Scienze dei Materiali, Piazza dell'ateneo nuovo 1, 20126, Milano, Italy

[d] AbAnalitica, Via Svizzera 16, 35027, Padova, Italy

[e] Università degli Studi di Genova, Dipartimento di Fisica, Via Dodecaneso 33, 16146, Genova, Italy

[f] Elements SRL, Viale Europa 596, 47521, Cesena, Italy

[g] The Institute of Scientific and Industrial Research, Osaka University, Mihogaoka 8-1, Ibaraki, Osaka 567-0047, Japan

[h] Dipartimento di Ingegneria Industriale, Università di Roma Tor Vergata, Via del Politecnico 1, 00133 Roma RM, Italia

[i] Department of Chemistry, Carl von Ossietzky University of Oldenburg, Oldenburg D-26129, Germany

[l] Dip. di Scienze e Metodi dell'Ingegneria, Università di Modena e Reggio Emilia, via Amendola 2, 42122 Reggio Emilia, Italy

[±] The authors equally contributed

*Corresponding Author Email: denis.garoli@unimore.it

**Abstract**

Ionic Current Rectification (ICR) can appear in nanopores, causing a diode-like behavior that originates from different efficiency of ion transport through the pore channel with respect to the applied voltage bias polarity. This effect is particularly interesting for nanopores with a short channel length, that is smaller than 500 nm, because then the dependence of the ion current on the direction along which the ions pass through the channel is determined by the geometry and material of the nanopore, and by the ion concentration in the electrolyte. Surface charges can play an important role, and therefore nanopores consisting of multiple materials can induce ICR because of different charge distributions at the nanopore-electrolyte interface for the different material sections. In this work, we study four solid-state nanopore designs considering different geometries of charge distributions along the inner and outer surfaces of nanopores, and evaluate their impact on ICR with six different electrolytes. The direct comparison between experimental data and modeling enables to understand how the different surface charge configurations impact the ICR.

1. **Introduction**

ICR[1-3] is a fundamental electrokinetic effect where a voltage of one polarity generates an ionic current that is notably higher than the one generated by the voltage of opposite polarity. Several possible trends and effects of ICR were studied, by means of nanoporous or membrane-based devices. In the last 30 years, biological[4,5], organic, inorganic[6] and organic-inorganic combined nanopores[1] were used to investigate ICR. In this context, from nanopores with diameters of few nanometers to micropores, a variety of diameters acted as probes for different applications, such as ionic rectification-enabling devices for ionic selectivity[7], ionic circuits and components[8], molecule-sensing[9], energy conversion[10] and desalination[11].

ICR occurs essentially due to the distribution of ions inside the channel of a solid-state nanopore or in its proximity, which strongly depends on the polarity voltage that is applied to the nanopore device. One voltage polarity causes an increase in the number of ions translocating through the pore (resulting in a high-conducting state), while the opposite voltage polarity creates a depletion of ion concentration in the nanopore (leading to a low-conducting state).[2,12] The origin of this behavior is structural or chemical asymmetry in the nanopore system. For example, one way to induce ICR in nanopores is by acting on their geometry, from cylindrical to conical[2]. Another approach is the introduction of gradients, including wettability[7], surface charge density[13], pH[14], viscosity[15], concentration[16] and charge distribution in the outer surfaces of the nanopore[17]. In addition to geometric asymmetry and gradients, other structural characteristics (such as pore length and channel diameter) play a fundamental role for the ICR. Early experiments in the field were performed on nanopores with long channels (exceeding 500 nm)[2,28]. However, reducing the length of the nanopore channel significantly boosts its bio-sensing abilities[2,18], therefore studies about ICR in short nanopores (i.e., <500 nm) have flourished. The majority of the short-channel nanopores that allowed ICR were fabricated with two different materials that constituted a bipolar junction. In this approach, the nanopore's inner walls would have two regions with opposite charge polarity, i.e., opposite ionic selectivity, granting ICR[2,12]. In this paper, we focused our attention on conical nanopores that had a channel length of 100 nm and a orifice diameter (i.e., the diameter at the small edge) below 100 nm. These nanopores were made of layers of different materials, i.e., $Si_3N_4$; $Si_3N_4//SiO_2$, $Si_3N_4//Ti//Au$ and $Si_3N_4//Ti//Au//SiO_2$ (see Section **2.Materials and Methods** for details). The rationale behind this choice was to study the effects of the junctions between the different materials on the ICR properties of each nanopore. Furthermore, we studied the ICR behaviour of each nanopore with six different electrolytes and we discussed their impact on the nanopores' ICR.

2. **Materials and methods**

***Materials. Solutions and Suspensions.*** In the rectification experiments, all the electrolyte solutions, i.e. $KCH_3COO$, $KBr$, $KCl$, $KNO_3$, $KClO_4$, $KPF_6$, were prepared at 10 mM concentration, using ultrapure MilliQ water, with resistivity of 18.2 MΩ·cm, as a solvent. All the solutions were prepared and used at room temperature (RT) and successively filtered with 200 nm-mesh Sartorius® filters. The same preparation process was used in the rectification experiment where different concentrations of KCl were tested (1 mM, 10 mM, 100 mM and 1000 mM). 10 mM electrolyte solutions were prepared at pH=7 and at pH=3. In each electrolyte solution at pH=3, the acid that corresponded to each solution's salt was used.

Iso-Propyl Alcohol (i.e., IPA 100% vol), MilliQ $H_2O$, acetic acid ($CH_3COOH$ 99,85% vol), hydrobromic acid (i.e., HBr 48% vol), hydrochloric acid (HCl 37% vol), nitric acid ($HNO_3$ 65% vol) and perchloric acid ($HClO_4$ 70% vol), were purchased from Sigma-Aldrich and were analytical grade or purer.

**Methods.** *Chip fabrication.* The structure of the four chips differed in their surface coating as follows: one series was fabricated without additional deposited films (i.e., bare $Si_3N_4$), and others with $SiO_2$, Ti//Au, or Ti//Au//$SiO_2$ coating layers. The thicknesses of the Ti, Au and $SiO_2$ coatings were, respectively: 3 nm, 20 nm and 5 nm. The depositions of Ti and of Au were performed by electron beam evaporation at a rate of 0.1 nm/sec. Subsequently, the chip was treated by Focused Ion Beam (FIB), to drill a (conical) hole into the membrane, constituting the nanopore. $SiO_2$ was deposited on the nanopore by means of plasma ALD at 110°C, with a final thickness of 5 nm, using BTBAS (i.e., bis(*tert*-butylamino)silane) as precursor.

*Electrical characterization.* Electrical measurements were performed by means of the e-NanoPore Reader (e-NPR) by Elements srl[29,30]. A frequency of 1.25 KHz was selected for the data acquisition. Ag/AgCl quasi-reference electrodes were used for these measures, and they were prepared by immersion of two Ag wires (0.5 mm diameter, 99.9% trace metals basis by Sigma

Aldrich) in 5% sodium-hypochlorite (Sigma Aldrich). The experimental protocol used to collect all the nanopores' ICR measures was a staircase-shaped cyclic voltammetry, where voltage was changed by 100 mV every 20 s from 1 V to -1 V and reverse. This meant registering the equivalents of reduction and oxidation branches of a cyclic voltammetry measure, according to IUPAC convention[26]. When applying a voltage for 20 s, the related electric current was the average of the values registered in the last 1 s of voltage application, in order to prevent oscillations[12]. To perform the measures, we used the electrode that was facing the bottom side of the nanopore as our anode (i.e., - electrode) and our reference electrode, while we used the electrode that was facing the orifice (i.e., tip side) of the nanopore as our cathode (i.e., + electrode) and as our working electrode.

*Ion beam and Electron beam characterization.* Nanopores were milled into the membrane using the $Ga^{+2}$ ions (Focused Ion Beam, with a current of 24 pA) produced in the Helios Nanolab 650 FEI workstation. The same equipment was used to inspect the nanopore's size with Scanning Electron Microscopy.

*Nanopore wetting procedure.* When a nanopore is used as a probe for single-entity detection, its proper wetting plays a fundamental role in the process. Indeed, if the nanopore were not wetted, the majority of its volume would be occupied by air bubbles, which would hinder both ionic and single-entity transport through the nanopore itself and through its feeding microfluidic circuit in general[34,35]. In addition, bubbles' translocation through the nanopore could result in fake electric translocation signals (i.e., fake current rectification, fake current drops, fake current blockades). For this reason, we designed our rectification experiment with an efficient wetting procedure composed of two parts: plasma treatment and nanopore wetting. As regards the plasma treatment, the whole assembled fluidic cell, with the nanopore chip inside, was exposed to 5 min of O$_2$ plasma at a power of 100 W. Then, the fluidic cell and nanopore assembly was wetted with a wetting solution that consisted in a 50:50%vol mixture of iso-propyl alcohol (IPA) and water. Three injections of wetting solution were performed before each experiment, letting the solution in contact with the microfluidic circuit 5 minutes every time. Then two injections of the measured electrolyte were performed. The first to wash the wetting solution and the second to perform the measurement. We found this wetting procedure granted reliable and effective ICR measurements.

*Models and numerical simulations.* The models and the numerical simulations were developed in COMSOL Multiphysics version 5.3. All simulations were provided using an axisymmetric system. A full and detailed description of simulations can be found in the Supporting Information. To describe the behavior of ionic current occurs in nanopore in electrolyte solutions Poisson-Nernst-Planck (PNP) (Eq. 1 and 2) and Navier-Stokes (NS) (Eq. 3 and 4) equations[37,38] were used. These equations govern the behavior of the electric potential on the entire domain (Poisson equation, Electrostatics module), ion flux in the electrolyte solution (Nernst-Planck, Transport of Diluted species module, fluid domain only) and fluid flow (Navier-Stokes, Laminar Flow, fluid domain only) in the simulating system.

$$\vec{J_i} = -D_i \nabla c_i + \frac{z_i F}{RT} D_i c_i \nabla \varphi + c_i \vec{u} \qquad (1)$$

$$\nabla^2 \varphi = -\frac{F}{\varepsilon} \sum_i z_i c_i \qquad (2)$$

$$\rho \left( \frac{\partial \vec{u}}{\partial t} + (\vec{u} \cdot \nabla)\vec{u} \right) = -\nabla p + \mu \nabla^2 \vec{u} - F \nabla \varphi \sum_i z_i c_i \qquad (3)$$

$$\nabla \cdot \vec{u} = 0 \tag{4}$$

In these equations $\vec{J_i}$ is the ion flux, $D_i$ is the diffusion coefficient, $c_i$ is the ion concentration, $z_i$ is the ion valency, $F$ is the Faraday constant, $R$ is the universal gas constant, T is the temperature, $\varphi$ is the potential, $\rho$ is the fluid density (was set as $1000\ kg \cdot m^{-3}$), $\mu$ is the viscosity of electrolyte solution (denotes as $0.001\ Pa \cdot s$), $\vec{u}$ is the fluid velocity, $p$ is the pressure.

To initiate electrical properties of gold a surface charge and a floating potential boundary condition (initial values ($Q_0 = 0\ C, V_{init} = 0\ V$) within the Electrostatics module were applied - check the **SI** for all the details). In contrast to silicon nitride, gold exhibits strong polarizability and its potential depends on the surrounding bias. By specifying a floating potential on a conductive material unconnected to the bias, it is possible to induce a polarizability of the metal as a reaction to the electrical field surrounding it (thereby to create charges of opposite polarity in response to the charges of surrounding substances)[12,39,40].

3. **Results and discussion**

*Introductive considerations about the rectification behavior of four nanopores made of different materials.*

As it was reported in literature, the ICR is strongly influenced by the ionic electrolyte and the nanopore shape, size and material of the nanopore that is analyzed[24,31]. For this reason, we focused our attention on the rectification behavior of a particular structure: conical nanopores with a diameter in the range of 70-94 nm in a 100 nm thick freestanding $Si_3N_4$ membrane. To investigate the effect of different anions on the ICR, we tested in six different ionic electrolytes, i.e. $KCH_3COO$; $KBr$; $KCl$; $KNO_3$ ; $KClO_4$ ; $KPF_6$, [23-25] at 10 mM salt concentration with the nanopores with different materials described above. The reason for the choice of these electrolytes was that, in the order that was specified above, they are part of the Hofmeister series[23-25], which is a well-known method for ion classification. The Hofmeister series has held since the late 1800s and it is based on the ability of the ions to salt-in and salt-out proteins[23-25]. In the last decades the Hofmeister series was proved also in biological and in solid state nanopores[23,24]. In this series, the behaviour of ions can change from kosmotropic (i.e., ordered interactions with the water network of hydrogen bonds) to chaotropic (i.e., unordered interactions with the water network of hydrogen bonds). With the change from kosmotropic to chaotropic, also the ICR curves of the electrolyte were reported to change from positive (i.e., exponential-like behaviour in the positive voltages region) to negative (i.e., logarithmic-like behaviour in the positive voltages region). The anions of $KCH_3COO$, $KBr$, $KCl$ and $KNO_3$ are kosmotoropic (i.e., positively rectifying) ions, while those of $KClO_4$ and $KPF_6$ are chaotropic (i.e., negatively rectifying) ions[23,24]. For this reason, even if our nanopores were not functionalised to enhance specific surface interactions with those ions, our tests aimed at studying the behaviour of the bare nanopores (i.e., the basic experimental conditions) with regard to their interactions with rectifying ions.

### *$Si_3N_4$ and $Si_3N_4//SiO_2$ nanopores.*

The $Si_3N_4$ and the $Si_3N_4//SiO_2$ nanopores were first verified in terms of pore diameter using a well-established method that is based on their conductance, which was estimated from the linear part of the experimental measures in KCl (i.e., [-300;+300] mV)[32]. The nanopores' tip diameters resulted to be about 90 nm. The first ICR experiment was performed with the bare $Si_3N_4$ nanopores. In Fig. 1A, experimental results showed that the shape of the I-V curve was similar for all the electrolytes, thus indicating that the nanopore's material, size and shape were the principal causes of its ICR behavior. Fig. 1B reported numerical results of the same system. In these simulations different types of the surface charge distribution on the nanopore were used as it was shown in Fig. 2,S2-S5 (all charged surface (ACS), a charged inner surface (ICS), a charged inner surface and exterior surface on the tip side (IECS)). In Fig. 1B the ACS model was used. The I-V curves were quantitatively similar to the experiments, confirming that the behavior of I-V curve can be mainly ascribed to the pore's shape and size and to the homogeneous negative surface charge distribution on the entire $Si_3N_4$ nanopore's surface. One of the advantages of simulations is that different distributions of surface charge can be explored allowing to shed light on the role of surface charge distribution on the I-V curve shape. Simulation with surface charge only on the top and the inner surface (IECS distribution) provides results (Supplementary Fig. S7) that are quite similar to the system with the ACS model. Instead, for the case with the ICS model, the I-V curve was qualitatively different (supplementary Fig. S8). This strongly indicated that charges on the top side (near the cone's tip) affected the ions fluxes much more than charges on the bottom (base aperture)[2].

The comparison between experiments and simulation allowed us to understand the possible role of the different anions. In the simulations, $KCH_3COOH$ (black) is indistinguishable from $KNO_3$ (green). This is expected because in the PNP-NS model, the only parameter used to describe the anions' motion is their diffusivity and the diffusivity of $CH_3COOH^-$ and $NO_3^-$ are very close, as reported in Supplementary Table S1. So, these two anions' I-V curves overlap. In the experiments, instead, the I-V curve of $KCH_3COOH$ (i.e., black in Fig. 1A,B) is quite different from the one of $KNO_3$ (i.e., green in Fig. 1A,B). This suggests that diffusivity is not the only relevant parameter ruling the ions' transport and that some additional phenomena may likely occur at the nanopore's walls. Similar arguments hold for KBr (i.e., red in Fig. 1A,B) and KCl (i.e., blue in Fig. 1A,B) that almost overlap in simulations while differing in experiments.

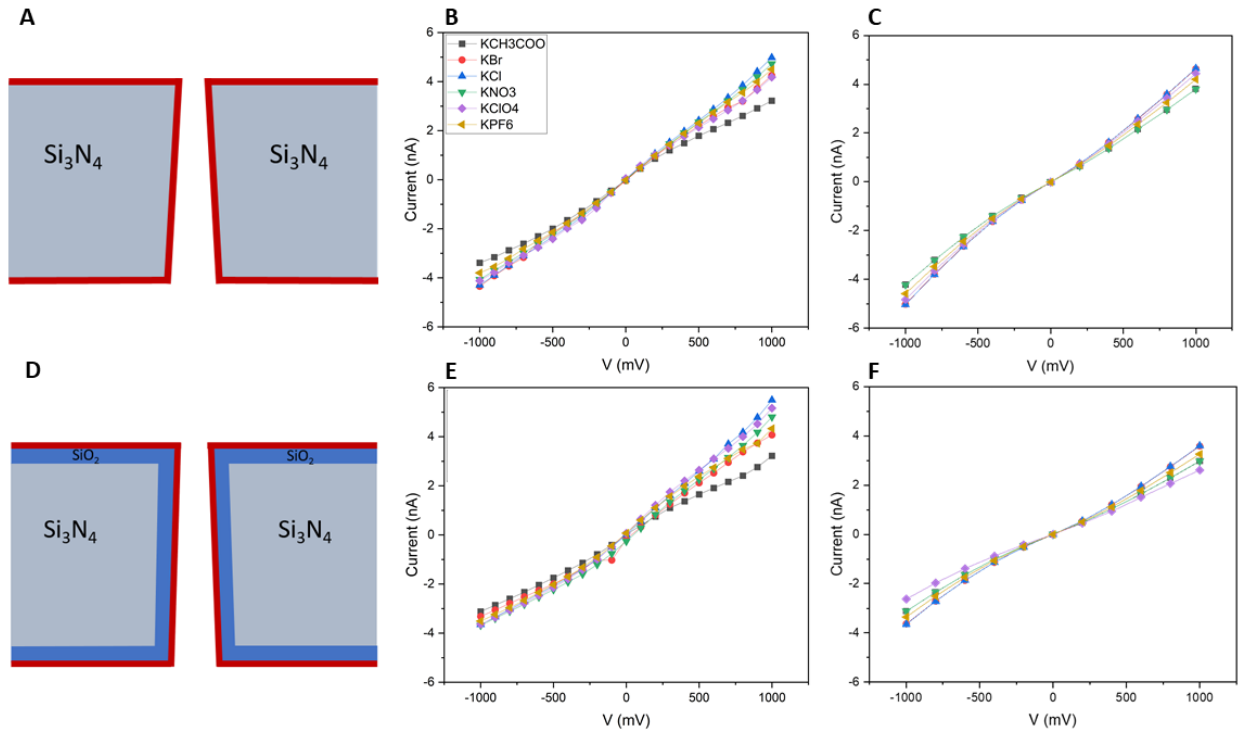

Figure 1 | Current rectification in $Si_3N_4$ and in $Si_3N_4//SiO_2$ nanopore. (A) Surface charge configuration for the $Si_3N_4$ nanopore. Surface charge distribution is represented by a red line on the nanopore's surfaces where charge is placed. (B) Experimental data, recorded in a single 94 nm $Si_3N_4$ nanopore. Six different electrolytes were tested in the same nanopore: $KCH_3COO$, KBr, KCl, $KNO_3$, $KClO_4$, $KPF_6$. The curves show a sigmoidal shape which is confirmed also by the simulations (C) that were performed assuming a homogeneous distribution of negative charge on the surface of the entire nanopore. (D) Surface charge configuration for the $Si_3N_4//SiO_2$ nanopore. Surface charge distribution is represented by a red line on the nanopore's surfaces where charge is placed. (E) The same experiment in (B) was performed on a 40 nm $Si_3N_4$ nanopore that was covered by a 5 nm-thick $SiO_2$ layer. Total diameter of the $Si_3N_4//SiO_2$ nanopore

was 94 nm. In both the experimental data (E) and the models (F), the presence of the $SiO_2$ layer made the curve hyperbolic.

The I-V curves only had minor changes when a $Si_3N_4$ nanopore was covered by a 5 nm-thick layer of $SiO_2$[44], still featuring a final diameter of about 90 nm, as it was reported in Fig. 1C. The I-V curves were slightly more asymmetric with respect to the previous case and featured current more intense at positive voltages than at negative ones.

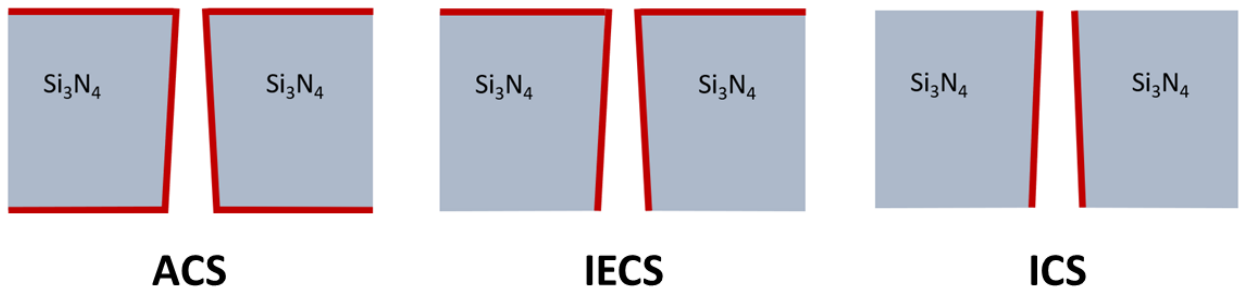

Figure 2. Examples of different surface charge configurations on a $Si_3N_4$ nanopore. All charged surface (ACS), a charged inner surface (ICS), a charged inner surface and exterior surface on the tip side (IECS).

This effect was better highlighted by the rectification ratios (i.e., the ratios of the currents' absolute values at the same positive and negative voltages, for example ±1V), that were reported in Fig 3B. Conversely, for the bare nanopore (i.e., $Si_3N_4$ in Fig. 2A), the logarithms of the rectification ratios of 2 electrolytes out of 6 were negative (i.e., the rectification ratios were lower than 1, or the absolute value of I was lower at +1V than at -1V). This was an interesting behavior, because the electrolytes in question were kosmotropic, so they were expected to have positive logarithms of the rectification ratio (i.e., logRR).

Thus, in our system, the behavior of these ions was different with respect to previous studies that were reported in literature[23] because our nanopores were not functionalized with specific molecules. Indeed, previous studies that were able to prove the Hofmeister series in solid-state nanopores regarded only polyimidazolium-brushes-functionalized glass nanopipettes, which had positive charged-groups exposed to the external environment and thus were prone to bind to the electrolyte's anions, i.e., the final object of the Hofmeister-series studies[23]. Therefore, without molecules to functionalize the nanopores' surface, it would be the bare nanopore's material (i.e., $Si_3N_4$) to interact with the electrolyte. This interaction would be characterized by factors such as the negative surface charge of the $Si_3N_4$, the charge of the ions, their ionic strength and the diameter of the nanopore channel. These four factors together could be the reasons for the negative rectification (ratios) of $KCH_3COO$ and $KBr$[45,46]. A key factor that allowed to produce the ICR measurements that we are presenting was the specific wetting procedure that we used to increase the nanopores' wettability. The oxygen plasma treatment would polish the surface of the treated materials on one hand, and on the other hand, it would introduce a certain number of dangling bonds on their surfaces. The following injection of IPA+$H_2O$ would exploit the low contact angle of IPA with the flow cell+nanopore assembly to distribute $H_2O$ on the surface of these materials, thus saturating most of the available sites for chemical bonds and exposing -OH groups on the treated surfaces. Finally, when water-based electrolytes would be injected in the microchannels, the exposed -OH groups would interact with water, and the overall treatment would result in an increased wettability of the nanopore.

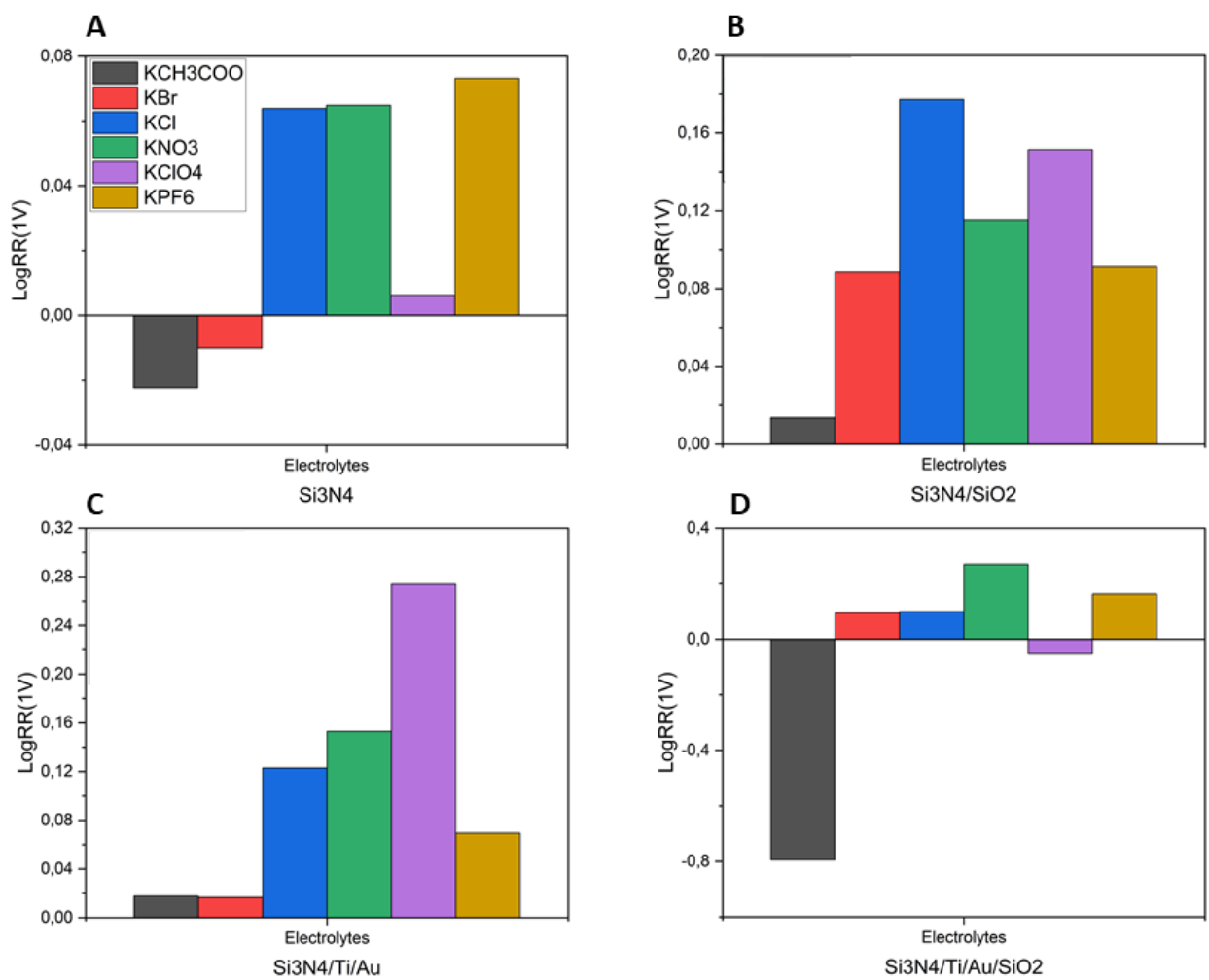

Figure 3 | LogRR(±1V) of the different nanopores. All the six electrolytes are considered for each nanopore. (A) logRR(±1V) of the bare $Si_3N_4$-nanopore chip. (B) LogRR(±1V) of the $Si_3N_4//SiO_2$-nanopore chip. (C) LogRR(±1V) of the $Si_3N_4//Ti//Au$-nanopore chip. (D) LogRR(±1V) of the $Si_3N_4//Ti//Au//SiO_2$-nanopore chip.

Finally, as we did for the $Si_3N_4$ nanopore, we performed simulations also for the $Si_3N_4 //SiO_2$ nanopores. The simulated currents (Fig. 1D) were slightly smaller than the ones in the $Si_3N_4$ nanopore (Fig. 1B). This was somehow expected since the $SiO_2$ coated pore may have had a smaller tip radius than the uncoated $Si_3N_4$ one. Also in this case, we repeated the simulations in different conditions (i.e., the ACS (Fig. 1D,2,S9), IECS (Fig. 2,S10) and ICS charge nanopores (Fig. 2,S11), highlighting, as in the case of $Si_3N_4$, the fundamental role of the surface properties on the nanopore's tip side and the minor role of surface modification at the larger bottom aperture of the nanopore. Even in these simulations, $KCH_3COOH$ (i.e., the black line in Fig. 1D) is indistinguishable from $KNO_3$ (i.e., the green line in Fig. 1D) as well as KBr (i.e., the red line in Fig. 1D) is indistinguishable from KCl (i.e., the blue line in Fig. 1D).

***$Si_3N_4$ nanopore characterization by changing the pH level of electrolytes solutions.***

In order to better understand the effect of pH on the rectification properties of our solid-state nanopores, we performed additional measurements on the $Si_3N_4$ nanopore with 5 of the 6 electrolytes that we used in the multi-electrolyte experiments. In the present case, we prepared all the electrolytes at pH=3, by the addition of the suitable volume of each electrolyte's corresponding acid. We tested all the electrolytes whose acid was available for us. As expected, the $Si_3N_4$ nanopore's ICR exhibited pH-dependent variability in response to different electrolyte solutions' pH values. This was due to protonation/deprotonation reactions occurring on the functional groups on the $Si_3N_4$ surface. Consequently, the change of surface charge density of the nanopore also took place. Experimental measurements, supported by theoretical calculations, were reported in Fig. 4.

Under acidic conditions at pH=3, the magnitude of positive current was observed to be smaller than the negative counterpart, in comparison with measurements that were made with a neutral pH=7 (Fig. 1). Experimental results aligned with theoretical predictions (see **SI** for the details of the pH calculations), except for the case of $KCH_3COO$, as we discuss at the end of the paragraph.

It is well known that the concentration of hydrogen ions depends on the pH level of the electrolyte solution, which affects the distribution of the surface charge of the nanopores[41,42]. The sign of the surface charge of the nanopore is tied to protonation/deprotonation reactions occurring within the amine and silanol functional groups of the $Si_3N_4$ nanopore[43]. Thus, under neutral pH conditions the negative surface charge is predominantly caused by the concentration of $OH^-$ ions, whereas at pH=3 it is caused by the concentration of $H^+$ ions. Typically, at neutral pH=7-8, the $Si_3N_4$ nanopore is negatively charged. When a positive voltage is applied, cations electrostatically accumulate near the walls of the nanopore. At pH=3 the surface charge density of the nanopore is supposed to be positive. In this case, anions are attracted towards the walls of the nanopores, while cations are repelled. Conversely, at neutral pH, under positive voltage, the walls of the nanopore are negatively charged and cations accumulate at the walls of the nanopore. Hence, at pH 3 the interaction between cations and the nanopore's positively charged walls is reduced, thus, the magnitude of ionic current under positive voltage is much less than under the negative voltage.

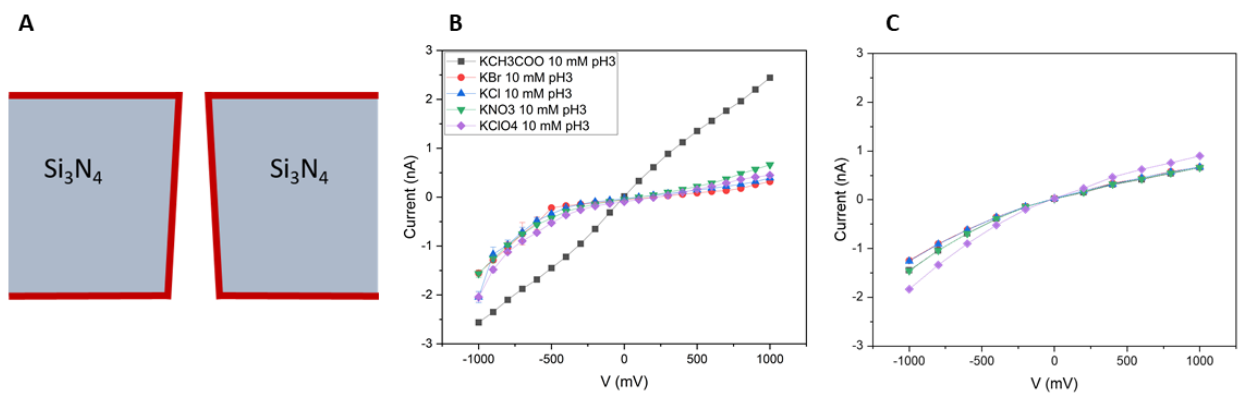

Figure 4| Rectification in $Si_3N_4$ nanopore. (A) Surface charge distribution of the studied nanopore (B) Experimental and (C) modeling data, recorded in a single $Si_3N_4$ nanopore. Five different electrolytes were

tested in the same nanopore: $KCH_3COO$, $KBr$, $KCl$, $KNO_3$, $KClO_4$. The rectification behavior of the nanopore is due to the change of the surface charge density of the nanopore at pH=3.

The observed ICR behavior was almost the same for all the electrolytes, with the exception of the potassium acetate (i.e., $KCH_3COO$). Potassium acetate dissociates into acetic acid (i.e., $CH_3COOH$), releasing a proton. Since acetic acid is a weak acid and has buffering properties, $KCH_3COO$ resists significant changes in pH level in the whole system[33]. For this reason, the curve corresponding to potassium acetate deviated from the others, showing minimal pH-related variations.

### *$Si_3N_4$//Ti//Au and $Si_3N_4$//Ti//Au//$SiO_2$ nanopores.*

It is known that Au coatings affect the rectification properties of nanopores, in the presence of different electrolytes[7,12,14]. In addition, Au-coated nanopores are attracting increasing research interest because of their potential plasmonic properties[19,20]. For example, these properties could lead to LSPR-mediated enhancement of the fluorescence emission of fluorophores, or nanoparticles, which could be used as label-markers for molecular structures in the context of nanopore-based memory technologies[19-22,36]. For these reasons, two nanopore structures that were studied in our experiments featured the base nanopore structure in $Si_3N_4$ with an adhesion layer of Ti (3 nm), that connected the top side of the nanopore to a layer of Au (20 nm-thick). Also in this case, the diameter of the $Si_3N_4$//Ti//Au was estimated by I-V based method[32] and resulted to be about 75 nm.

As reported in Fig. 5B, the gold layer that coated the top side of the nanopore was responsible for a significant change in its ICR curves. Indeed, the ICR curves of this nanopore featured a clear sigmoidal behaviour, with opposite convexities and higher current intensities than the bare $Si_3N_4$ nanopore. As in the previous cases, we investigated the reasons for the particular shape of the ICR curves by means of numerical simulations and only considering the presence of gold and its surface charge distribution, it has been possible to resemble the experimental data (Fig 5C).

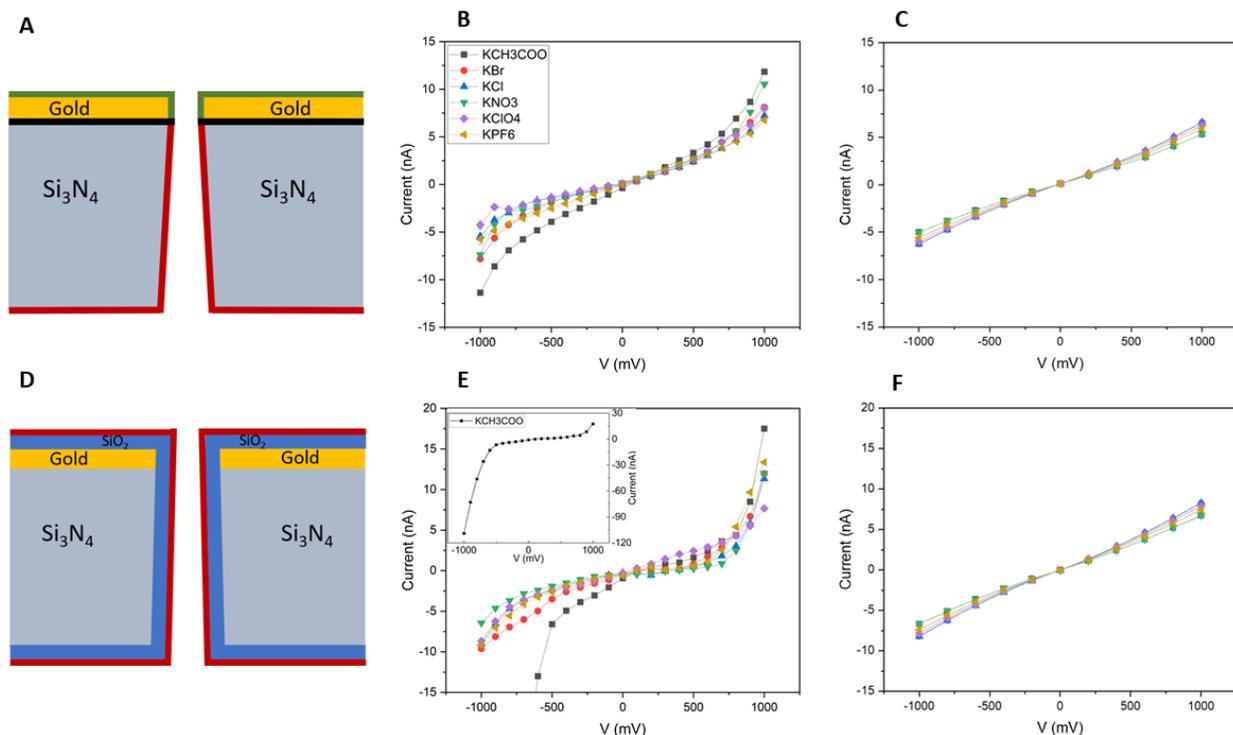

Figure 5 | Rectification in $Si_3N_4$//Ti//Au and in $Si_3N_4$//Ti//Au//$SiO_2$ nanopore. (A) Surface charge configuration for the $Si_3N_4$//Ti//Au nanopore. Surface charge distribution is represented by a red line on the nanopore's surfaces where charge is placed. (B) Experimental data, recorded in a single 40 nm $Si_3N_4$/Ti/Au nanopore. The Ti layer was 3 nm thick and the Au layer was 20 nm thick. Six different electrolytes were tested in the same nanopore: $KCH_3COO$, KBr, KCl, $KNO_3$, $KClO_4$, $KPF_6$. The curves show a sigmoidal shape which is confirmed also by the modeling (C). The sigmoidal shape has a different concavity and the current has a higher intensity with respect to the cases in Fig. 1A. (D) Surface charge configuration for the $Si_3N_4$//Ti//Au//$SiO_2$ nanopore. Surface charge distribution is represented by a red line on the nanopore's surfaces where charge is placed. (E-F) The same experiment in (B-C) was performed on a 40 nm $Si_3N_4$/Ti/Au/$SiO_2$ nanopore that was covered by a 5 nm-thick $SiO_2$ layer. In this case, the rectification behavior of the nanopore is dominated by the presence of $SiO_2$ on its surface.

When an extra layer of SiO$_2$ was added on top of the Si$_3$N$_4$//Ti//Au nanopore, the ICR curve remained sigmoidal and almost all the rectification ratios were preserved. However, the key features of the ICR curves, in this case, were not due to the presence of Au in this system as well, but most likely to the surface charge present on the top SiO$_2$ layer that faced the electrolyte. Indeed, simulations were performed with the ACS, IECS and ICS$^2$ charged SiO$_2$ layer (Fig. 6,S12, S13, S14) and only the case that is represented in Fig. 5D could partially explain what we observed experimentally. It was the particular surface charge distribution of the SiO$_2$ (or Au) top layer and its interaction with the ions in the electrolyte that determined the shape of the ICR curves (see Fig. S12-S17 in **SI** for different charge configurations on the Au or SiO$_2$ surface and their ICR). In Fig. 6,7 the ICS and IECS surface charge configuration of the Si$_3$N$_4$//Ti//Au and Si$_3$N$_4$//Ti//Au//SiO$_2$ nanopore were reported. Here is represented that also in these cases, the highest contribute to the rectification behavior of the nanopore was given by the surfaces on the nanopore's tip and inner side, while the ICR contribution of the bottom surfaces of the nanopore was less marked and it mainly caused the node-like behavior in the [0; 300] mV range.

The ICR trends of the Si$_3$N$_4$//Ti//Au and Si$_3$N$_4$//Ti//Au//SiO$_2$ nanopores were also pictured in Fig. 3C and Fig. 3D, where the logRR of the two nanopores were reported with respect to each electrolyte. All the logRR were positive in the Au surface-case, while only the one for KClO$_4$ was negative in the case of the Au//SiO$_2$ nanopore. As already mentioned in the previous paragraph, this was an effect of the size, material, surface charge of the nanopores and electrolyte environment.

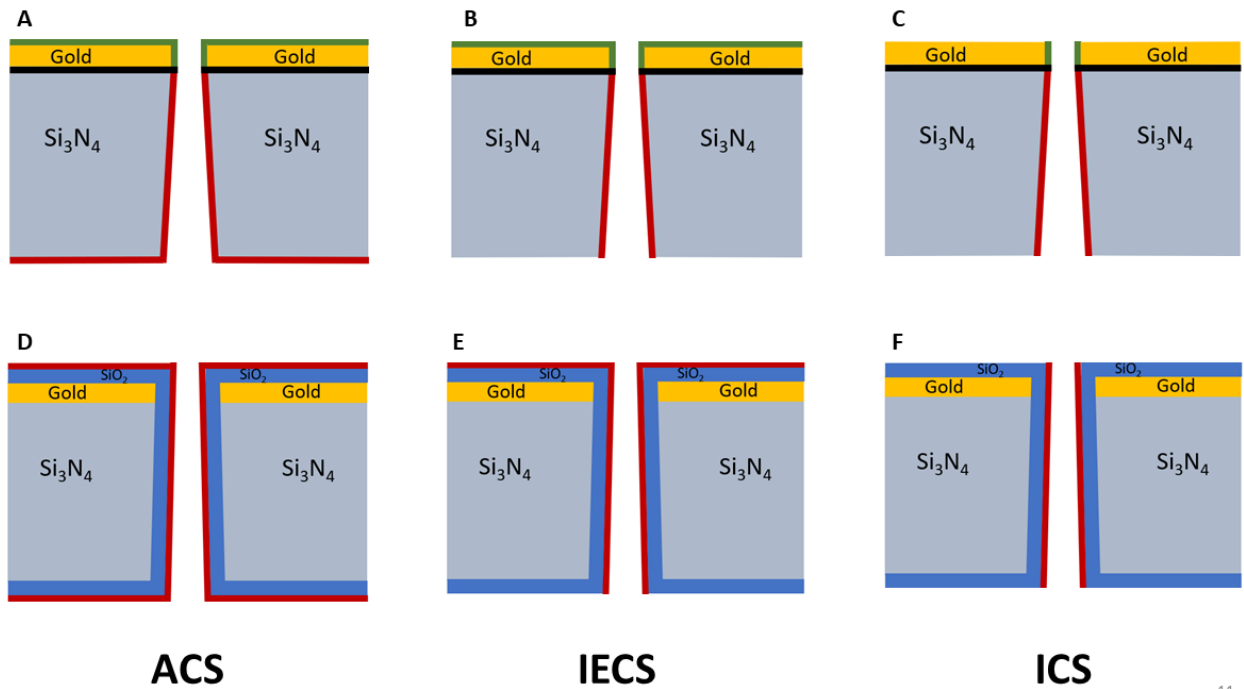

**Figure 6** | ACS, IECS and ICS surface charge distribution of the $Si_3N_4//Ti//Au$ and $Si_3N_4//Ti//Au//SiO_2$ nanopore. Red and green lines on the different surfaces represent the regions where surface charge is placed. On the red lines, surface charge that is simulated according to standard COMSOL methods, while on green lines according to floating potential method, as explained in Section 2 of the paper.

4. **Conclusions**

In this paper, we reported about the study of the ICR characteristics of four types of nanopores, made of layers of different materials: $Si_3N_4$; $Si_3N_4//SiO_2$, $Si_3N_4//Ti//Au$ and $Si_3N_4//Ti//Au//SiO_2$. We fixed the nanopores' length (i.e., 100 nm), tip diameter (i.e., 76-95 nm) and shape profile and we studied how hydrophilic ($SiO_2$) and/or plasmonic materials (Au) layers modified their performance[19-22]. In addition, we studied the effect of different electrolytes on the ICR behaviour of the different nanopores. Their behaviour varied sensibly in each case, in terms of shape of the curve, absolute current intensities and rectification ratio. As it was widely reported in literature, and in agreement with our numeric simulations, the basic reason of these phenomena lies in the charge distribution at the surfaces of the nanopores[2]. Indeed, different charge distribution and polarity on the surfaces of the nanopore cause different interactions between the nanopore and the translocating anions. Thus, specifically modifying the ability of the anions to translocate the nanopore and thus their ICR curve.

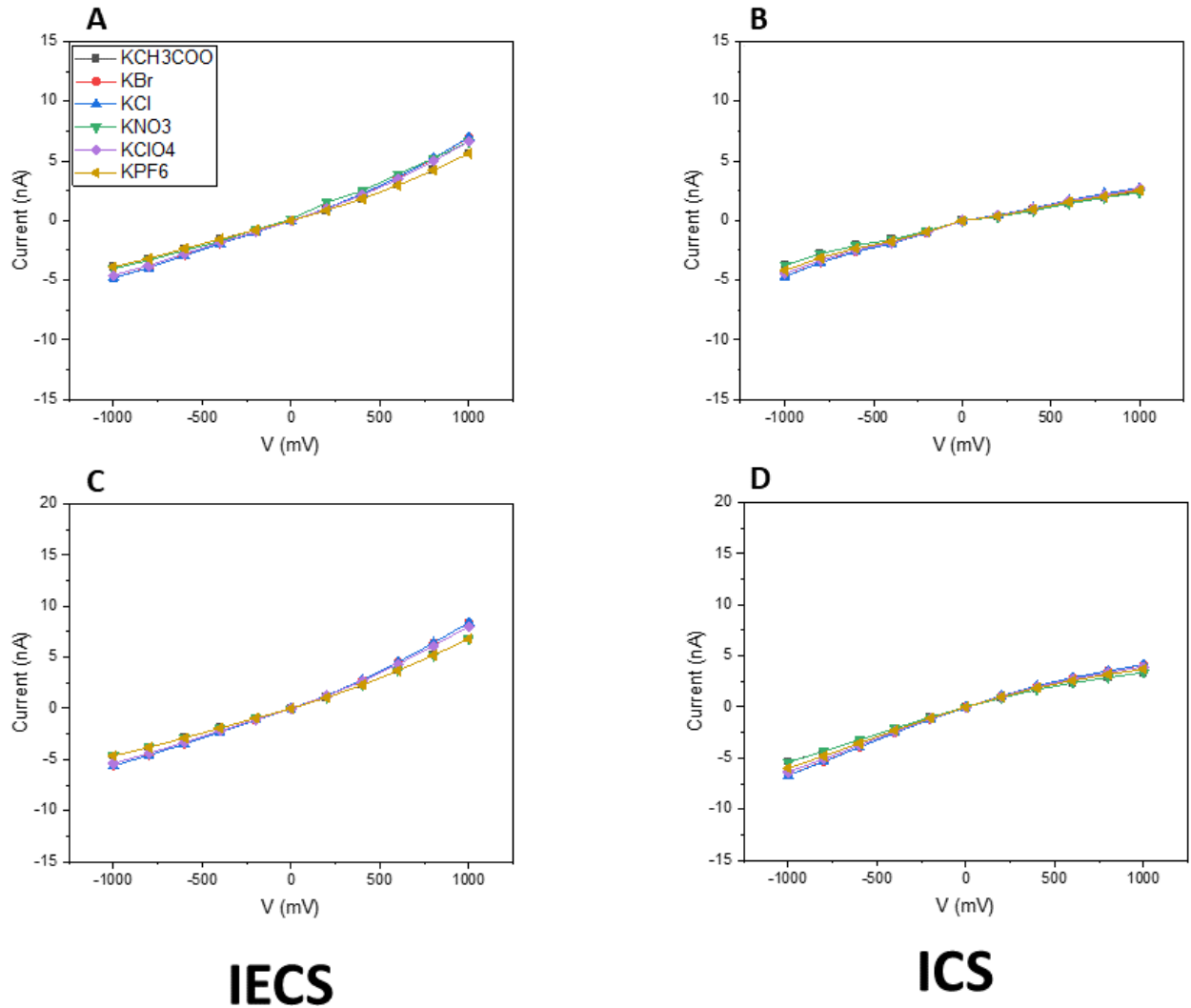

Figure 7 | ICR with IECS and ICS surface charge distribution of the $Si_3N_4//Ti//Au$ (A,B) and $Si_3N_4//Ti//Au//SiO_2$ (C,D) nanopore. IECS surface charge configuration contribution to the overall ICR behaviour was more prominent than the one of the ICS surface charge configuration, as it was represented in Fig 5B,E.

Therefore, in addition to ions, all charged entities (i.e., dsDNA, ssDNA, single molecules, nanoparticles) could experience different electrostatic interactions while translocating nanopores, according to the (concentration of the) electrolytic buffer they will be dispersed into and to the nanopore itself. Thus, different electric potentials, currents and translocation profiles (i.e., current blockade shapes) will characterize the different cases.

To sum up, in this paper we provided a comprehensive overview of the basic properties of four different kinds of nanopores in regard of ICR, considering a set of electrolytes used to study the well-known Hofmeister series. We reported and discussed the behavior of these nanopores and the effects that fabrication material, size, shape and surrounding electrolyte have on ICR. With the aim of building an essential tool for the nanopore scientific community, we provided the definition of specific pre-treatments that can be applied to the nanopores to enhance the quality and reliability of the measures. We think that the analyses and the procedures that we reported in this paper could constitute an essential tool for any scientist who is approaching the field of solid-state nanopores sensors.

**Acknowledgements**

The authors thank the European Union under the Horizon 2020 Program, FET-Open: DNA-FAIRYLIGHTS, Grant Agreement 964995 and the HORIZON-MSCA-DN-202: DYNAMO, grant

Agreement 101072818, D.M. acknowledges the financial support from the European Research Council (ERC) under the European Union's Horizon 2020 research and innovation program (Grant Agreement 948238).